\documentclass[twocolumn,prc,amssymb, aps,superscriptaddress, 
amsmath]{revtex4}

\usepackage{color}
\usepackage{amsmath,mathtools,booktabs,
            microtype} %,siunitx}

\usepackage{graphicx}
\usepackage{dcolumn}
\usepackage{bm}
\usepackage{multirow}
\usepackage{times}
\usepackage{url}
\usepackage{tikz}
\usepackage[version=4]{mhchem}

\usetikzlibrary{arrows,decorations.pathmorphing}
\usetikzlibrary{arrows.meta}
\usetikzlibrary{calc}

\definecolor{violet}{HTML}{602969}
\definecolor{red}{HTML}{FC0009}
\definecolor{orange}{HTML}{FF6319}
\definecolor{green}{HTML}{00933C}
\definecolor{blue}{HTML}{0036A6}
\definecolor{yellow}{HTML}{FFBE00}
\definecolor{lightgrey}{HTML}{A7A9AC}

\newcommand{\be}{\begin{equation}}
\newcommand{\ee}{  \end{equation}}
\newcommand{\ba}{\begin{eqnarray}}
\newcommand{\ea}{  \end{eqnarray}}
\newcommand{\ve}{\varepsilon}

\begin{document}

\title{Transition-State Theory Revisited}

\author{Hans A. \surname{Weidenm\"uller}}
\email{haw@mpi-hd.mpg.de}
\affiliation{Max-Planck-Institut f\"ur Kernphysik, Saupfercheckweg 1,
  D-69117 Heidelberg, Germany}

\date{\today}

%%%%%%%%%%%%%%%%%%%%%%%%%%%%%%%%%%%%%%%%%555

\begin{abstract}Two quantum systems, each described as a
  random-matrix ensemble. are coupled to each other via a number of
  transition states. Each system is strongly coupled to a large number
  of channels. The average transmission probability is the product of
  three factors describing, respectively, formation of the first
  system from the entrance channel, decay of the second system through
  the exit channel, and transport through the transition states. Each
  of the transition states contributes a Breit-Wigner resonance. In
  general, the resonances overlap.

\end{abstract}
%%%%%%%%%%%%%%%%%%%%%%%%%%%%%%%%%%%%%%%%%%%%%%

\maketitle

%%%%%%%%%%%%%%%%%%%%%%%%%%%%%%%%%%%%%%%%%%%%%%%%%%%%%%%%%%%%%%%%%%%%%%%%

\section{Introduction}

Starting with the seminal work of Bohr and Wheeler~\cite{Boh39},
transition-state theory has played an important role in applications
of many-body quantum theory to physics and chemistry, see, for
instance, Refs.~\cite{Hae90, Tru96}. The theory aims at calculating
the probability of passage through or over a barrier that separates
two parts of a physical many-body system, and of the associated
reaction rates. Following a suggestion in Refs.~\cite{Hag21, Ber21},
an approach from first principles to transition-state theory has been
developed in Ref.~\cite{Wei22}. The approach describes the two parts
of the system in terms of two independent random-matrix
ensembles. Each of these is coupled to a number of channels that feed
or deplete the system. In Ref.~\cite{Wei22} the model was worked out
explicitly for quantum tunneling through a barrier, and for passage
through a single transition state above the barrier (which turns into
a transition-state resonance). The model has been generalized in
Ref.~\cite{Hag23} to the situation where the two parts of the system
are coupled by a number of non-overlapping transition-state
resonances, and where both parts are strongly coupled to a large
number of channels. The present note describes a further
generalization of that model by allowing the transition-state
resonances to overlap. Such generalizations may apply, for instance,
to the transmission of electrons through quantum dots~\cite{Alh00}.

\section{Model}
\label{mod}

We consider two time-reversal invariant quantum systems, each coupled
to a set of scattering states. The two systems are coupled to each
other via a set of $k$ transition states. In matrix form, the
Hamiltonian is
\ba
\label{d1}
H = \left( \begin{matrix} H_1 & V_1 & 0 \cr
  V_1^T & H_{\rm tr} & V_2^T \cr
  0 & V_2 & H_2 \cr \end{matrix} \right) \ .
\ea
Here $H_1$ ($H_2$) is the real and symmetric Hamiltonian matrix
governing system $1$ (system $2$, respectively). It acts in Hilbert
space $1$ (in Hilbert space $2$, respectively). Both Hilbert spaces
have dimension $N$. The real and symmetric Hamiltonian matrix $H_{\rm
  tr}$ acts in the transition space which has dimension $k$. The
coupling matrices $V_1$ and $V_2$ are real and have $N$ rows and $k$
columns each. The upper index $T$ denotes the transpose.

We assume that the matrices $H_1$ and $H_2$ are statistically
independent members of the Gaussian Orthogonal Ensemble (GOE) of
random matrices. Their elements are zero-centered real Gaussian random
variables with second moments
\ba
\label{d2}
\langle (H_1)_{\mu_1 \mu'_1} (H_1)_{\mu_2 \mu'_2} \rangle &=& \frac{\lambda^2}{N}
(\delta_{\mu_1 \mu_2} \delta_{\mu'_1 \mu'_2} + \delta_{\mu_1 \mu'_2} \delta_{\mu'_1
  \mu_2}) \ , \nonumber \\
\langle (H_2)_{\nu_1 \nu'_1} (H_2)_{\nu_2 \nu'_2} \rangle &=& \frac{\lambda^2}{N}
(\delta_{\nu_1 \nu_2} \delta_{\nu'_1 \nu'_2} + \delta_{\nu_1 \nu'_2} \delta_{\nu'_1
  \nu_2}) \ .
\ea
The angular brackets denote the ensemble average. The parameter
$\lambda$ defines the ranges of the two spectra. All indices labeled
$\mu, \mu'$ run from $1$ to $N$, those labeled $\nu, \nu'$ from $N + k
+1$ to $2 N + k$. Elements of the matrix $H_{\rm tr}$ carry indices
$m, m', n, n'$ that range from $N + 1$ to $N + k$. We eventually
consider the limit $N \to \infty$, keeping $k$ fixed. We assume that
the $k$ eigenvalues $E_m$ of $H_{\rm tr}$ are all located near the
centers of the spectra of $H_1$ and $H_2$.

We show in the Appendix that we may always write $V_1$ and $V_2$ in
the form
\ba
\label{d4}
(V_1)_{\mu m} &=& \sum_{m'}(O_1)_{\mu m'} {\cal V}_{1, m'}
(O_{\rm tr, 1})_{m m'} \ , \nonumber \\
(V_2)_{\nu m} &=& \sum_{m'} (O_2)_{\nu m'} {\cal V}_{2, m'}
(O_{\rm tr, 2})_{m m'} \ .
\ea
Here $O_1$ and $O_2$ are orthogonal matrices of dimension $N$, and
$O_{\rm tr, 1}$ and $O_{\rm tr, 2}$ are orthogonal matrices of
dimension $k$. Eqs.~(\ref{d4}) show that the matrices $V_1$ and $V_2$
are linear superpositions of the $k$ vectors $Z_{1, m'}$ and $Z_{2,
  m'}$ in transition space, with components
\ba
\label{d4b}
(Z_{1, m'})_m &=& {\cal V}_{1, m'} (O_{\rm tr, 1})_{m m'} \ ,
\nonumber \\
(Z_{2, m'})_m &=& {\cal V}_{2, m'} (O_{\rm tr, 2})_{m m'} \ ,
\ea
and with lengths $| {\cal V}_{1 m'} |$ and $| {\cal V}_{2 m'} |$,
respectively. The arguments in Ref.~\cite{Wei22} show that for all
$m'$, $|{\cal V}_{1, m'}|$ and $|{\cal V}_{2, m'}|$ must be small
compared to $\lambda$.

Each of the two GOE Hamiltonians is coupled to a set of open channels
labeled, respectively $a, a', a''$ etc. for $H_1$ and $b, b', b''$
etc. for $H_2$. The real coupling matrix elements are labeled $W_{a
  \mu}$ and $W_{b \nu}$, respectively. These obey
\ba
\label{d4a}
\sum_\mu W_{a \mu} W_{a' \mu} = \delta_{a a'} v^2_a \ , \ \sum_\nu W_{b \nu}
W_{b' \nu} = \delta_{b b'} v^2_b \ .
\ea
The relations~(\ref{d4a}) define the coupling strengths $v^2_a$ and
$v^2_b$. The Kronecker deltas rule out direct (i.e., non-resonant)
scattering processes $a \to a'$ and $b \to b'$~\cite{Ver85}. We define
the $N \times N$ width matrices $\Gamma_1$ and $\Gamma_2$ as
\ba
\label{d5}
(\Gamma_1)_{\mu \mu'} &=& 2 \pi \sum_a W_{a \mu} W_{a \mu'} \ , \nonumber \\
\ (\Gamma_2)_{\nu \nu'} &=& 2 \pi \sum_b W_{b \nu} W_{b \nu'} \ ,
\ea
and, in analogy to Eq.~(\ref{d1}), the total width matrix $\Gamma$ as
\ba
\label{d6}
\Gamma = \left( \begin{matrix} \Gamma_1 & 0 & 0 \cr
                                      0 & 0 & 0 \cr
                               0 & 0 & \Gamma_2 \cr \end{matrix} \right) \ . 
\ea
With $E$ the total energy of the system, the inverse of the propagator
matrix of the system is
\ba
\label{d7}
D(E) = {\bf E} - H + (i/2) \Gamma \ .
\ea
Here and in what follows, ${\bf E}$ stands for the product of the
variable $E$ and the unit matrix in the space under consideration.

There are two types of scattering processes. (i) Backscattering from
channel $a$ to channel $a'$ (or from channel $b$ to channel $b'$) on
the same side of the transition region, and (ii) transmission from
channel~$a$ through the transition region to channel~$b$ or vice
versa. The elements of the scattering matrix $S(E)$ for backscattering
are given by
\ba
\label{d8}
S_{a a'}(E) &=& \delta_{a a'} - 2 i \pi \sum_{\mu \mu'} W_{a \mu}
(D^{- 1}(E))_{\mu \mu'} W_{a' \mu'} \ , \nonumber \\
S_{b b'}(E) &=& \delta_{b b'} - 2 i \pi \sum_{\nu \nu'} W_{b \nu}
(D^{- 1}(E))_{\nu \nu'} W_{b' \nu'} \ ,
\ea
those for transmission through the transition region are given by
\ba
\label{d9}
S_{a b}(E) &=& - 2 i \pi \sum_{\mu \nu} W_{a \mu} (D^{- 1}(E))_{\mu \nu}
W_{b \nu} \ , \nonumber \\
S_{b a}(E) &=& - 2 i \pi \sum_{\nu \mu} W_{b \nu} (D^{- 1}(E))_{\nu \mu}
W_{a \mu} \ .
\ea
It is easy to check that the $S$-matrix is unitary and symmetric.

\section{Many Strongly Coupled Channels}
\label{man}

We calculate $\langle S_{a b}(E) \rangle$ and $\langle | S_{a b}(E)
|^2 \rangle$ by averaging over both $H_1$ and $H_2$.  We expand $(D^{-
  1}(E))_{\mu \nu}$ in powers of $V_1$ and $V_2$. We resum the
resulting series. To write the result in compact form, we define the
Green functions
\ba
\label{a1}
G_1(E) &=& ({\bf E} - H_1 + (i/2) \Gamma_1)^{- 1} \ , \nonumber \\
G_2(E) &=& ({\bf E} - H_2 + (i/2) \Gamma_2)^{- 1} \ , \nonumber \\
G_{\rm tr}(E) &=& ({\bf E} - H_{\rm tr} - V^T_1 G_1(E) V_1 \nonumber \\
&& \qquad \ \ \  - V^T_2 G_2(E) V_2)^{- 1} \ .
\ea
With these definitions, the first of Eqs.~(\ref{d9}) becomes
\ba
\label{a2}
&& S_{a b}(E) = - 2 i \pi \nonumber \\
&& \times \sum_{\mu \nu} W_{a \mu} \bigg( G_1(E) V_1
G_{\rm tr}(E) V^T_2 G_2(E) \bigg)_{\mu \nu} W_{b \nu} \ .
\ea
Eqs.~(\ref{a1}) show that $G_1(E)$, $G_2(E)$, and $G_{\rm tr}(E)$ are
statistically correlated. Therefore, it is not possible, in general,
to calculate $\langle S_{a b}(E) \rangle$ and $\langle | S_{a b}(E)
|^2 \rangle$ analytically. Complete analytical expressions can be
obtained, however, if the numbers of channels strongly coupled to
$H_1$ and to $H_2$ are both large compared to unity. That is the case
we consider. To quantify that condition we consider backscattering for
a closed transition space defined by putting $V_1 = 0 = V_2$ in
Eq.~(\ref{d1}). Eqs.~(\ref{d8}) then define two backscattering
matrices $S_1(E)$ and $S_2(E)$, given by
\ba
\label{m0}
(S_1(E))_{a a'} &=& \delta_{a a'} - 2 i \pi \sum_{\mu \mu'} W_{a \mu}
G_1(E)_{\mu \mu'} W_{a' \mu'} \ , \nonumber \\
(S_2(E))_{b b'} &=& \delta_{b b'} - 2 i \pi \sum_{\nu \nu'} W_{b \nu}
G_2(E)_{\nu \nu'} W_{b' \nu'} \ .
\ea
These depend, respectively, only on $H_1$ (only on $H_2$). The
ensemble averages of $(S_1(E))_{a a'}$ and $(S_2(E))_{b b'}$ are
diagonal. That follows~\cite{Ver85} from Eqs.~(\ref{d4a}). The average
values define the transmission coefficients
\ba
\label{m1}
T_a = 1 - |\langle (S_1)_{a a} \rangle|^2 \ , \ T_b = 1 -
|\langle (S_2)_{b b} \rangle|^2 \ .
\ea
We assume $\sum_a T_a \gg 1$, $\sum_b T_b \gg 1$. The strong
inequalities imply that terms of order $1 / \sum_a T_a$ and $1 /
\sum_b T_b$ are negligible compared to terms of order unity. We
calculate the leading-order terms in an asymptotic expansion of
$\langle S_{a b}(E) \rangle$ and of $\langle | S_{a b}(E) |^2 \rangle$
in inverse powers of $\sum_a T_a$ and $\sum_a T_b$.

For $\sum_a T_a \gg 1$, $\sum_b T_b \gg 1$, the arguments in
Ref.~\cite{Wei22} show that in the expression for $G_{\rm tr}(E)$ in
Eq.~(\ref{a1}) we may replace $G_1(E)$ and $G_2(E)$ by their average
values. With $E$ close to the centers of the spectra of $H_1$ and
$H_2$, these averages are given by
\ba
\label{m2}
\langle G_1(E) \rangle_{\mu \mu'} &=& - (i / \lambda) \delta_{\mu \mu'}
\ , \nonumber \\
\langle G_2(E) \rangle_{\nu \nu'} &=& - (i / \lambda) \delta_{\nu \nu'} \ .
\ea
Insertion into the last Eq.~(\ref{a1}) and use of the
definitions~(\ref{d4b}) gives
\ba
\label{m3}
G_{\rm tr}(E) &=& \bigg( {\bf E} - H_{\rm tr} + i V^T_1 V_1 / \lambda 
+ i V^T_2 V_2 / \lambda \bigg)^{- 1} \nonumber \\
&=& \bigg( {\bf E} - H_{\rm tr} + i \sum_{j = 1}^2 \sum_m Z^T_{j, m} 
Z_{j, m} / \lambda \bigg)^{- 1} \ .
\ea
We note that Eq.~(\ref{m3}) does not contain any random
elements. Substituting $G_{\rm tr}(E)$ given by Eq.~(\ref{m3}) into
Eq.~(\ref{a2}) yields an expression for $S_{a b}(E)$ that depends upon
$H_1$ ($H_2)$ only via the factor $G_1(E)$ (the factor $G_2(E)$,
respectively). That allows us to calculate $\langle S_{a b}(E)
\rangle$ and $\langle | S_{a b}(E) |^2 \rangle$ analytically.

\section{Average Transmission Amplitude}
\label{ave}

We calculate $\langle S_{a b}(E) \rangle$ by averaging separately the
factors that depend upon $H_1$ and on $H_2$. For the first of these we
use Eqs.~(\ref{d4}) and the definition of the vectors $Z_{1, m'}$ in
Eqs.~(\ref{d4b}) to write
\ba
\label{a3}
&& \sum_\mu W_{a \mu} (G_1(E) V_1)_{\mu m} \nonumber \\
&& \ \ = \sum_{\mu \mu' m'} W_{a \mu} (G_1(E))_{\mu \mu'} (O_1)_{\mu' m'}
(Z_{1, m'})_m \ . 
\ea
We define
\ba
\label{a4}
\tilde{H}_1 &=& O^T_1 H_1 O_1 \ , \nonumber \\
\tilde{W}_{a \mu} &=& \sum_{\mu'} W_{a \mu'} (O_1)_{\mu' \mu} \ ,
\nonumber \\
\tilde{G}(E) &=& O^T_1 G_1(E) O_1 \ .
\ea
Then expression~(\ref{a3}) takes the form
\ba
\label{a5}
\sum_{\mu m'} \tilde{W}_{a \mu} (\tilde{G}_1(E))_{\mu m'} (Z_{1, m'})_m \ . 
\ea
We observe that with $H_1$, the matrix $\tilde{H}_1$ is also a member
of the GOE, and that the coefficients $\tilde{W}_{a \mu}$ also obey
the first of Eqs.~(\ref{d4a}). We use the first of Eqs.~(\ref{m2}) for
$\tilde{G}_1(E)$. For expression~(\ref{a5}) that gives
\ba
\label{a5a}
- \frac{i}{\lambda} \sum_{m'} \tilde{W}_{a m'} (Z_{1, m'})_m \ .
\ea
The sum over $m'$ has $k$ terms. The first of Eqs.~(\ref{d4a}) shows
that for $N \to \infty$, $\tilde{W}_{a \mu}$ is inversely proportional
to $1 / \sqrt{N}$.  It follows that for fixed $k$ and $N \to \infty$,
expression~(\ref{a5a}) vanishes. Thus,
\ba
\label{a6}
\langle S_{a b}(E) \rangle = 0 \ .
\ea
The same conclusion is reached when we consider, instead of
expression~(\ref{a3}), the term $\sum_\nu (V^T_2 G_2(E))_{m \nu} W_{b
  \nu}$ in Eq.~(\ref{a2}).

\section{Average Transmission Probability}
\label{tra}

With $\langle S_{a b}(E) \rangle = 0$, the average probability for
transmission from channel $a$ to channel $b$ is given by
\ba
\label{a7}
P_{a b}(E) = \langle | S_{a b}(E) |^2 \rangle \ .
\ea
The average is over both $H_1$ and $H_2$. We use Eq.~(\ref{m3}) in
Eq.~(\ref{a2}). We use the transformation leading to Eq.~(\ref{a5})
and the corresponding transformation for the term containing
$G_2(E)$. We calculate
\ba
\label{m4}
X_{a, m' m''} &=& 2 \pi \bigg\langle \sum_\mu \tilde{W}_{a \mu}
\tilde{G}_1(E)_{\mu m'} \sum_{\mu'} \tilde{W}_{a \mu'}
\tilde{G}^*_1(E)_{\mu' m''} \bigg\rangle \ , \nonumber \\
X_{b, m' m''} &=& 2 \pi \bigg\langle \sum_\nu \tilde{G}_2(E)_{m' \nu}
\tilde{W}_{b \nu} \sum_{\nu'} \tilde{G}^*_2(E)_{m'' \nu'}
\tilde{W}_{b \nu'} \bigg\rangle \ . \nonumber \\
\ea
The first (second) angular bracket denotes the average over $H_1$
(over $H_2$, respectively). A slight generalization of the argument
used in the Appendix of Ref.~\cite{Wei22} shows that in the limit of
many strongly coupled channels we have
\ba
\label{m5}
X_{a, m' m''} &=& \delta_{m' m''} \frac{1}{\lambda} \frac{T_a}{\sum_{a'}
  T_{a'}} \ , \nonumber \\
X_{b, m' m''} &=& \delta_{m' m''} \frac{1}{\lambda} \frac{T_b}{\sum_{b'}
  T_{b'}} \ .
\ea
Using Eqs.~(\ref{m5}) in the expression for $| \langle S_{a b} \rangle
|^2$, we obtain for the average tranmission probability $P_{a b}$
defined in Eq.~(\ref{a7}) the expression
\ba
\label{m6}
P_{a b}(E) = \frac{T_a}{\sum_{a'} T_{a'}} Y \frac{T_b}{\sum_{b'} T_{b'}}
\ea
where
\ba
\label{m7}
Y = \sum_{m n} \bigg| \sum_{m' n'} (z_{1, m})_{m'} (G_{\rm tr}(E))_{m' n'}
(z_{2, n})_{n'} \bigg|^2 \ ,
\ea
where $G_{\rm tr}(E))$ is given by Eq.~(\ref{m3}), and where
\ba
\label{m6a}
z_{1, m} = Z_{1, m} / \sqrt{\lambda} \ , \ z_{2, m} = Z_{2, m} /
\sqrt{\lambda} \ .
\ea
Eq.~(\ref{m6}) gives the average transmission probability as the
product of three factors. The first (last) factor is the probability
to enter (leave) space~$1$ (space~$2$) via channel~$a$ (via
channel~$b$, respectively). The fact that $P_{a b}(E)$ factorizes and
the forms of the first and last factors are all due to the statistical
properties of $H_1$ and $H_2$, and to the assumptions $\sum_a T_a \gg
1$, $\sum_b T_b \gg 1$. That is seen by comparing Eq.~(\ref{m6}) with
the result of the theory of compound-nucleus
scattering~\cite{Ver85}. There one uses a scattering matrix of the
form of $(S_1(E))_{a a'}$ as in Eqs.~(\ref{m0}), with $H_1$ a member
of the GOE. For $a \neq a'$, for $N \to \infty$, and for $\sum_{a''}
T_{a''} \gg 1$, the resulting expression for $\langle | (S_1(E))_{a
  a'} |^2 \rangle$ factorizes and has the form~\cite{Ver86}
\ba
\label{m7a}
\langle | (S_1(E))_{a a'} |^2 \rangle = T_a \frac{T_{a'}}{\sum_{a''}
  T_{a''}} \ , \ a \neq a' \ .
\ea
That form is known as ``independence of formation and decay of the
compound nucleus'', with $T_a$ interpreted as the probability of
compound-nucleus formation, and the second factor as the relative
probability of compound-nucleus decay into channel
$a'$. Eq.~(\ref{m6}) differs in form from Eq.~(\ref{m7a}) in that it
contains two factors each denoting a relative probability. That is a
consequence of the symmetry of $S_{a b}(E)$ in Eqs.~(\ref{d9}) under
the operation $a \leftrightarrow b$.

The factor $Y$ in Eq.~(\ref{m7}) describes transport through the
transition space. It does not depend on the couplings of system~$1$
and system~$2$ to the channels. We show presently that this apparent
independence is the result of the assumptions $\sum_a T_a \gg 1$ and
$\sum_b T_b \gg 1$. The factor $Y$ is the sum of squares of amplitudes
each of which bears a close formal analogy to the inelastic parts of
the $S$-matrices in Eqs.~(\ref{m0}). Indeed, the transition space is
entered from space~$1$ through ``channel''~$m$ and left for space~$2$
through ``channel''~$n$ via the vectors $z_{1, m}$ and $z_{2, n}$,
respectively. The elements $(z_{1, m})_{m'}$ and $(z_{2, n})_{m'}$ are
analogues of the amplitudes $\sqrt{2 \pi} W_{a \mu}$ and $\sqrt{2 \pi}
W_{b \nu}$ in Section~\ref{mod}. All have dimension
(energy)$^{1/2}$. Propagation within the transition space is governed
by the effective Hamiltonian
\ba
\label{m8}
{\cal H}_{\rm eff} = H_{\rm tr} - i \sum_{j = 1}^2 \sum_m z^T_{j, m} z_{j, m}
\ea
which differs from $H_{\rm tr}$ because it accounts for the coupling
to spaces~$1$ and $2$. The coupling term has the same form and plays
the same role as the width matrices in Eqs.~(\ref{d5}, \ref{d7}). It
differs from the width term in Eq.~(\ref{d7}) by a factor two. While
the width matrix in Eq.~(\ref{d7}) is due to the free propagator in
channel space, the coupling term in Eq.~(\ref{m8}) is a remnant of the
propagators $G_1(E)$ and $G_2(E)$ in spaces~$1$ and $2$,
respectively. Propagation in these spaces is subject to the coupling
to the channels. In the limits $\sum_a T_a \gg 1$ and $\sum_b T_b \gg
1$, $G_1(E)$ and $G_2(E)$ take the forms~(\ref{m2}) wherein all
explicit reference to the channels disappears. Nevertheless, the
coupling to the channels leaves us with the above-mentioned factor
two.

The Hamiltonian ${\cal H}_{\rm eff}$ is complex symmetric and can be
diagonalized by a complex orthogonal matrix ${\cal O}$,
\ba
\label{m9}
({\cal H}_{\rm eff})_{m n} = \sum_{l = 1}^k {\cal O}_{m l} {\cal E}_l
  {\cal O}_{n l} \ .
\ea
The $k$ complex eigenvalues ${\cal E}_l$, $l = 1, \ldots, k$ obey $\Im
({\cal E}_l) < 0$ for all $l$. For $j = 1, 2$ we define amplitudes
\ba
\label{m10}
\zeta_{j, m l} = \sum_{m'} (z_{j, m})_{m'} {\cal O}_{m' l} 
\ea
and obtain
\ba
\label{m11}
Y = \sum_{m n} \bigg| \sum_l \zeta_{1, m l} \frac{1}{E - {\cal E}_l}
\zeta_{2, n l} \bigg|^2 \ .
\ea
Inserting that into Eq.~(\ref{m6}) gives our final result,
\ba
\label{m12}
&& P_{a b}(E) \\
&& = \frac{T_a}{\sum_{a'} T_{a'}} \sum_{m n} \bigg| \sum_l
\zeta_{1, m l} \frac{1}{E - {\cal E}_l} \zeta_{2, n l} \bigg|^2
\frac{T_b}{\sum_{b'} T_{b'}} \ . \nonumber
\ea
The total transmission probability is given by
\ba
\label{m12a}
\sum_b P_{a b}(E) = \frac{T_a}{\sum_{a'} T_{a'}} \sum_{m n} \bigg| \sum_l
\zeta_{1, m l} \frac{1}{E - {\cal E}_l} \zeta_{2, n l} \bigg|^2 \ .
\ea
In Eqs.~(\ref{m12}, \ref{m12a}), each transition state contributes to
transmission through the transition space a Breit-Wigner resonance
with complex resonance energy ${\cal E}_l = \ve_l - i \gamma_l$. The
resonance energies are ordered by putting $\ve_1 < \ve_2 < \ldots <
\ve_k$. Eqs.~(\ref{m2}) may be used in $G_{\rm tr}$ only if all $\ve_l$
are close to the centers $E = 0$ of the spectra of $H_1$ and $H_2$. An
equivalent assumption was formulated in Section~\ref{mod}.

To interpret Eqs.~(\ref{m12}, \ref{m12a}), we consider two extreme
cases. (i) The $k$ resonances are isolated so that $\ve_l - \ve_{l -
  1} >> \gamma_l, \gamma_{l - 1}$ for all $l = 2, \ldots, k$. Then $Y$
reduces to
\ba
\label{m13}
Y = \sum_{m n l} \bigg| \zeta_{1, m l} \frac{1}{E - {\cal E}_l}
\zeta_{2, n l} \bigg|^2 \ .
\ea
Transmission is described by a sum of Lorentzians. That is somewhat
similar to the case considered in Ref.~\cite{Hag23} where simplifying
statistical assumptions on the matrices $V_1$ and $V_2$ are
used. These prevent the resonances from overlapping and cause all
resonances to have the same partial and total widths. (ii) All
resonances overlap so that $\gamma_l, \gamma_{l - 1} \leq \ve_l -
\ve_{l - 1}$ for all $l$. Then
\ba
\label{m14}
Y = \sum_{m n} \sum_{l l'} \zeta_{1, m l} \frac{1}{E - {\cal E}_l}
\zeta_{2, n l} \zeta^*_{1, m l'} \frac{1}{E - {\cal E}^*_{l'}}
\zeta^*_{2, n l'}  \ ,
\ea
with nonnegligible interference terms between pairs $l \neq l'$ of
resonances. The values of $Y$ for cases (i) and (ii) may differ
substantially. The actual physical situation may lie anywhere between
cases~(i) and (ii). It seems that only case~(i) has so far been
considered in the literature. Presence or absence of interference
terms could be tested experimentally using a beam with variable mean
energy $E$ and band width $\delta E < \ve_l - \ve_{l - 1}$.

{\bf Acknowledgement.} The author is grateful to K. Hagino for
correspondence.

\section*{Appendix}

We derive the first of Eqs.~(\ref{d4}), the second follows
analogously. The matrices $V_1 V^T_1$ and $V^T_1 V_1$ have,
respectively, dimension $N$ and $k$. Both are real and
symmetric. Thus, $V_1 V^T_1$ and $V^T_1 V_1$ are diagonalized by
orthogonal matrices $O_1$ of dimension $N$ and and $O_{\rm tr, 1}$ of
dimension $k$, respectively, so that
\ba
\label{ap1}
(V_1 V^T_1)_{\mu \mu'} &=& \sum_{s = 1}^k (O_1)_{\mu s} {\cal V}^2_s
(O_1)_{\mu' s} \ ,
\nonumber \\
(V^T_1 V_1)_{m m'} &=& \sum_{t = 1}^k (O_{\rm tr, 1})_{m t} {\cal W}^2_t
(O_{\rm tr, 1})_{m' t} \ . 
\ea
Both matrices $V_1 V^T_1$ and $V^T_1 V_1$ are positive semidefinite.
Therefore, all eigenvalues are positive or zero. In Eqs.~(\ref{ap1}),
that fact is taken into account by writing the eigenvalues as squares
of real numbers. The rank of $V_1$ is $k$. Therefore, at most $k$
eigenvalues of $V_1 V^T_1$ differ from zero. That fact is indicated in
the summation over $s$ in the first of Eqs.~(\ref{ap1}). In the
identity $V_1 (V^T_1 V_1) = (V_1 V^T_1) V_1$ we replace the contents
of the round brackets by the right-hand sides of
Eqs.~(\ref{ap1}). After some simple matrix algebra the identity takes
the form
\ba
\label{ap2}
      [O^T_1 V_1 O_{\rm tr, 1}]_{\mu m} {\cal W}^2_m = {\cal V}^2_\mu
      [O^T_1 V_1 O_{\rm tr, 1}]_{\mu m} \ .
\ea
Eq.~(\ref{ap2}) implies ${\cal W}^2_m = {\cal V}^2_m$ for all $m = 1,
\ldots, k$ and $[O^T_1 V_1 O_{\rm tr, 1}]_{\mu m} \propto \delta_{\mu
  m}$. Insertion of that relation into the first of Eqs.~(\ref{ap1})
shows that the proportionality constant is ${\cal V}_m$, so that
$[O^T_1 V_1 O_{\rm tr, 1}]_{\mu m} = {\cal V}_m \delta_{\mu m}$. That
yields the first of Eqs.~(\ref{d4}).

%%%%%%%%%%%%%%%%%%%%%%%%%%%%%%%%%

\end{document}